\documentstyle[graphicx,prl,aps]{revtex}
\begin{document}
	
\twocolumn[                              
\hsize\textwidth\columnwidth\hsize\csname@twocolumnfalse\endcsname  

\title{Wave Function of a Brane-like Universe}
\author{Aharon Davidson, David Karasik, and Yoav Lederer}
\address{Physics Department,
Ben-Gurion University of the Negev,
Beer-Sheva 84105, Israel\\
(e-mail: davidson@bgumail.bgu.ac.il)}
\maketitle

\begin{abstract}
	Within the mini-superspace model, brane-like cosmology
	means performing the variation with respect to the
	embedding (Minkowski) time $\tau$ before fixing the
	cosmic (Einstein) time $t$.
	The departure from Einstein limit is parameterized by
	the 'energy' conjugate to $\tau$, and characterized
	by a classically disconnected Embryonic epoch.
	In contrast with canonical quantum gravity, the
	wave-function of the Universe is (i) $\tau$-dependent,
	and (ii) vanishes at the Big Bang.
	Hartle-Hawking and Linde proposals dictate discrete
	'energy' levels, whereas Vilenkin proposal resembles
	$\alpha$-disintegration.
\end{abstract}
\pacs{PACS numbers: }]    

In this paper, the Universe is viewed as a curved
four-dimensional bubble\cite{bubble} floating in a higher
dimensional flat background.
To discuss the quantum disintegration of such a Brane-like
Universe and derive the corresponding time-dependent and
Big-Bang resistant wave function, we restrict ourselves to
the framework of the mini-superspace model\cite{mini}.

Assuming the Universe to be homogeneous, isotropic, and
closed, the Friedman-Robertson-Walker (FRW) line element
can be written in the form
\begin{equation}
	ds^{2} = \sigma^{2}\left[-({\dot \tau}^{2}-
	{\dot a}^{2})dt^{2}+
	a^{2}(t)d\Omega_{3}^{2}\right] ~,
\end{equation}
where $d\Omega_{3}^{2}$ denotes the metric of a unit
$3$-sphere, and $\sigma^{2}\equiv \frac{4G}{3\pi}$ is
a normalization factor.
One may still exercise of course the gauge freedom of
fixing $\tau(t)$. 
However, the more general form helps us keep track of
the way the FRW manifold is embedded within a $5$-dim
Minkowski spacetime
\begin{equation}
	ds^{2}_{5}=\sigma^{2}\left[-d\tau^{2}+da^{2}+
	a^{2}d\Omega_{3}^{2}\right] ~.
	\label{5flat}
\end{equation}
A pedagogical case of sufficient complexity involves a
positive cosmological constant $\Lambda$.
The corresponding mini-Lagrangian, defined by
$I=\int_{}^{}{\cal L}dt$, is given by
\begin{equation}
	{\cal L} = -\left(\frac{a{\dot a}^{2}}
	{\sqrt{{\dot \tau}^{2}-{\dot a}^{2}}} +
	a(H^{2}a^{2}-1)\sqrt{{\dot \tau}^{2} -
	{\dot a}^{2}}\right) ~,
\end{equation}
where $H^{2}\equiv\frac{16\pi G}{3} \Lambda$.

$\bullet$ The \textit{standard} mini-superspace
prescription is to impose the so-called cosmic gauge,
namely ${\dot \tau}^{2}-{\dot a}^{2}=1$, and treat
$a(t)$ as a single canonical variable.
This way, the variation with respect to $a(t)$ gives rise
to Raychaudhuri equation, and the complementary evolution
equation
\begin{equation}
	{\dot a}^{2}+1 = H^{2}a^{2}
\end{equation}
is then nothing but the Arnowitt-Deser-Misner (ADM)
Hamiltonian constraint ${\cal H}=0$.
Let $P=-2a{\dot a}$ be the momentum conjugate to $a$,
the Hamiltonian takes the form
\begin{equation}
	{\cal H} = -\frac{1}{4a}\left(P^{2}+V(a)\right) ~,
\end{equation}
involving the familiar potential
\begin{equation}
	V(a) = 4a^{2}\left(1-H^{2}a^{2}\right) ~.
\end{equation}
Quantization means replacing $\displaystyle{P
\rightarrow -i\frac{\delta}{\delta a}}$ and imposing the
Wheeler-DeWitt (WDW) equation ${\cal H}\Psi(a)=0$ on the wave
function of the Universe.

$\bullet$ The \textit{non-standard} procedure would
be to allow both $a(t)$ and $\tau(t)$ to serve as two
independent canonical variables.
The variation with respect to $\tau(t)$ results in a
simple conservation law.
Owing to ${\cal L}(a,\dot a,\dot \tau)$, the 'energy'
$\omega$ conjugate to $\tau$ is conserved.
Imposing the cosmic gauge ${\dot \tau}^{2}-{\dot a}^{2}=1$
only at this stage, we can rearrange the 'energy' conservation
equation into a generalized evolution equation
\begin{equation}
	{\dot a}^{2}+1 = \xi H^{2}a^{2} ~,
	\label{RT}
\end{equation}
with $\xi(a)$ being a root of
\begin{equation}
	\xi (\xi-1)^{2}=\frac{\omega^{2}}{H^{6}a^{8}} ~.
	\label{xi}
\end{equation}
Eq.(\ref{RT}) is recognized as the Regge-Teitelboim 
(RT)\cite{RT,RTcosmo} equation of motion, with Einstein limit
approached as $\omega\rightarrow 0$, that is $\xi \rightarrow1$
(the physical brunch is identified with $\xi \geq 1$).
This comes with no surprise, given that RT canonical
variables are in fact the embedding coordinates.
Recalling that classical RT-cosmology\cite{RTcosmo} involves
only one independent equation of motion, the equation arising
by varying with respect to $a(t)$ is superfluous.

The $2$-momentum $\displaystyle{P_{\alpha}=(\frac{\delta
{\cal L}}{\delta{\dot \tau}},\frac{\delta {\cal L}}{\delta
{\dot a}})}$ is given by
\begin{mathletters}
\label{P}
\begin{eqnarray}
	P_{\tau} & = & 
	\left[\left(\frac{\dot a}{\sqrt{{\dot \tau}^{2} -
	{\dot a}^{2}}}\right)^{2} + 1 - H^{2}a^{2}\right]
	\frac{a\dot \tau}{\sqrt{{\dot \tau}^{2}-{\dot a}^{2}}} ~, \\
	P_{a} & = &
	-\left[\left(\frac{\dot a}{\sqrt{{\dot \tau}^{2} -
	{\dot a}^{2}}}\right)^{2} + 3 - H^{2}a^{2}\right]
	\frac{a\dot a}{\sqrt{{\dot \tau}^{2}-{\dot a}^{2}}} ~.
\end{eqnarray}
\end{mathletters}
The $t$-derivatives which enter eq.(\ref{P}) conveniently
furnish a time-like unit $2$-vector
\begin{equation}
	n^{\alpha} \equiv
	\left(\frac{\dot \tau}{\sqrt{{\dot \tau}^{2}-{\dot a}^{2}}},
	\frac{\dot a}{\sqrt{{\dot \tau}^{2}-{\dot a}^{2}}}\right)
	~,~~ n^{2}+1=0 ~.
\end{equation}
Invoking now a $(\dot \tau,\dot a)$-independent matrix
\begin{equation}
	\rho^{\alpha}_{\,\beta} =
	\left(
	\begin{array}{cc}
			2a(H^{2}a^{2}-1) & 0  \\
			0 & 2a(H^{2}a^{2}-2)
		\end{array}
	\right) ~,
	\label{rho}
\end{equation}
$P^{\alpha}$ can be put in the compact form
\begin{equation}
	P^{\alpha} = \frac{1}{2}(n\rho n)n^{\alpha} +
	\rho^{\alpha}_{\beta}n^{\beta} ~,
\end{equation}
after subtracting $a(H^{2}a^{2}-1)(n^{2}+1)n^{\alpha}$.
A naive attempt to solve $n^{\alpha}(\rho, P)$, as
apparently dictated by the Hamiltonian formalism,
and substitute into the constraint $n^{2}+1 = 0$,
falls short. 
The cubic equation involved does not admit a simple
solution, and the resulting constraint is anything
but quadratic in the momenta.

The way out, that is linearizing the problem, involves the
definition of an independent quantity $\lambda$, such that
\begin{equation}
	n\rho n +2\lambda = 0 ~.
\end{equation}
Off the Einstein limit, $\lambda$ is not an eigenvalue of
$\rho^{\alpha}_{\,\beta}$, and we can solve for $n^{\alpha}
(\rho,P,\lambda)$ to find
\begin{equation}
	n^{\alpha}=\left[\left(\rho-\lambda 
	I\right)^{-1}\right]^{\alpha}_{\,\beta}P^{\beta} ~.
\end{equation}
This allows us to finally convert the combined constraints
$n^{2}+1=0$ and $n\rho n +2\lambda-\lambda (n^{2}+1)=0$ into
\begin{equation}
	\left\{
	\begin{array}{c}
		P(\rho-\lambda I)^{-2}P + 1 = 0 ~,\\
		P(\rho-\lambda I)^{-1}P + \lambda = 0 ~.
		\end{array}
	\right.
	\label{PP}
\end{equation}
The first equation is the derivative with respect to
$\lambda$ of the other.
This suggests that $\lambda$ be elevated to the level
of a canonical non-dynamical variable in the forthcoming
Hamiltonian formalism.

Needless to say, the above seems to be a tip of a bigger
iceberg, a mini-superspace version of Brane-like gravity.
Indeed, carrying out the (say) $10$-dim RT embedding of
the $4$-dim ADM formalism, we have recently derived the
quadratic Hamiltonian\cite{RTH}
\begin{equation}
	{\cal H} = \frac{1}{2}N
	\left[P_{A}\left((\rho-\lambda I)^{-1}\right)^{AB}P_{B} +
	\lambda \right] + N^{i}y^{A}_{\,,i}P_{A} ~,
\end{equation}
where the novel Lagrange multiplier $\lambda$ accompanies the
standard non-dynamical variables, the lapse function $N$ and
the shift vector $N^{i}$.
To shed light on the matrix $\rho^{AB}$, one infers that
\begin{equation}
	\frac{\rho^{AB}}{ 2\sqrt{h}}=(h^{ia}h^{jb}-h^{ij}
	h^{ab})y^{A}_{\,|ab}y^{B}_{\,|ij}+\left({\cal R}^{(3)}+
	6H^{2}\right)\eta^{AB} ~,
\end{equation}
with ${\cal R}^{(3)}$ denoting the $3$-dim Ricci scalar
constructed by means of the spatial $3$-metric
$h_{ij}=\eta_{AB}y^{A}_{|i}y^{B}_{|j}$.

The quantum theory dictates ${\displaystyle P_{A}\rightarrow
-i\frac{\delta}{\delta y^{A}}}$.
The corresponding wave function $\Psi(\tau,a)$ is subject
to two Virasoro-type constraints.
The so-called momentum constraint equation $\displaystyle
{y^{A}_{,i}\frac{\delta\Psi}{\delta y^{A}}=0}$, which is
trivially satisfied at the mini-superspace level, is
accompanied by a \textit{bifurcated} WDW equation
\begin{equation}
	\left\{
	\begin{array}{c}
		\displaystyle{\frac{\delta}{\delta y^{A}}
		\left((\rho-\lambda I)^{-1}\right)^{AB}
		\frac{\delta}{\delta y^{B}}\Psi = \lambda\Psi} ~, \\
		\displaystyle{\frac{\delta}{\delta y^{A}}
		\left((\rho-\lambda I)^{-2}\right)^{AB}
		\frac{\delta}{\delta y^{B}}\Psi = \Psi} ~.
	\end{array}
	\right.
	\label{WDW}
\end{equation}
Given the diagonal $\rho^{\alpha}_{\,\beta}$ specified by
eq.(\ref{rho}), and up to all sorts of order ambiguities,
the mini-superspace wave function $\Psi(\tau,a)$ obeys
\begin{mathletters}
\begin{eqnarray}
	-\frac{\partial^{2}\Psi}{\partial \tau^{2}} & = &
	\xi(\xi-1)^{2}H^{6}a^{8}\Psi ~,
	\label{psitau} \\
	-\frac{\partial^{2}\Psi}{\partial a^{2}} & = &
	a^{2}\left[2+(\xi-1)H^{2}a^{2}\right]^{2}
	(-1+\xi H^{2}a^{2})\Psi ~,
	\label{psia}
\end{eqnarray}
\end{mathletters}
where the $\lambda \leftrightarrow \xi$ dictionary
reads
\begin{equation}
	\lambda \equiv a\left[(\xi+1)H^{2}a^{2}-2\right] ~.
\end{equation}
The separation of variables is accomplished by
substituting $\Psi(\tau,a)=\psi(a)\chi(\tau)$.
Eq.(\ref{psia}) then tells us that $\xi=\xi(a)$.
In turn, eq.(\ref{psitau}) can admit a solution only
provided $\xi(a)$ is such that $\xi(\xi-1)^{2}H^{6}a^{8}
=\omega^{2}$ is a constant.
This is how the conserved 'energy' $\omega$, introduced
by eq.(\ref{xi}), enters the quantum game.

Altogether, the $\tau$-dependent wave function of
the Universe acquires the familiar form
\begin{equation}
	\Psi(\tau,a) = \psi(a)e^{-i\omega\tau} ~,
\end{equation}
with the $\tau$-dependence dropping out at the Einstein limit.
The radial component $\psi(a)$ satisfies the residual WDW
equation
\begin{equation}
	\left(-\frac{\partial^{2}}{\partial a^{2}} +
	V(a)\right)\psi = 0 ~,
\end{equation}
where the modified potential, depicted in fig.(1),
is given explicitly by
\begin{equation}
	V(a) = a^{2}\left[2+(\xi-1)H^{2}a^{2}\right]^{2}
	(1-\xi H^{2}a^{2}) ~.
\end{equation}
\begin{figure}[tbp]
	\begin{center}
		\includegraphics[scale=0.5]{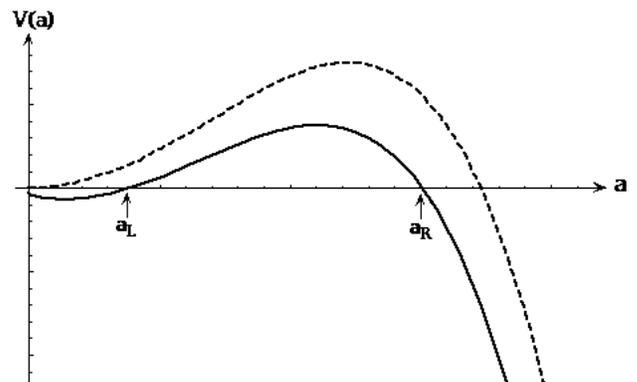}
		\caption{Wheeler-DeWitt potential for a Brane-like
		Universe (solid curve), and for Einstein Universe
		(dashed curve).}
		\label{fig1}
	\end{center}
\end{figure}
$V(a)$ admits a barrier provided $\omega H\leq\frac{2}
{3\sqrt{3}}$, which we now adopt as the case of interest.
The barrier is stretched between $a_{L}<a<a_{R}$, where
$a_{L,R}$ are the two positive roots of $H^{2}a^{3}-a+
\omega=0$.
For $\omega H\ll 1$, the classical turning points are
located at
\begin{equation}
	a_{L}\simeq \omega ~,~~
	a_{R}\simeq H^{-1}(1-\frac{1}{2}\omega H) ~.
\end{equation}	
At long distances, only a slight deviation from the
original potential is detected, namely
\begin{equation}
	V(a \gg \omega) \simeq
	4a^{2}(1-H\omega-H^{2}a^{2}) ~.
\end{equation}
But at short distances, a serendipitous well (with a
surplus of `kinetic energy' at the origin) makes its
appearance
\begin{equation}
	V(a \leq \omega) \simeq
	-\omega^{2} -3\omega^{4/3}a^{2/3} +4a^{2} ~.
	\label{well}
\end{equation}
The emerging classically disconnected Embryonic epoch
is the essence of brane-like quantum cosmology.

A theory of boundary conditions is still to be
constructed.
The situation is even more complicated in a scheme
where the Big-Bang is classically alive and cannot
be traded for a Euclidean conic-singularity-free pole.
The Riemann tensor gets pathological as $a\rightarrow 0$,
leaving us with no alternative but to interpret 'nothing'
\cite{nothing} as
\begin{equation}
	\Psi(\tau,a=0) = 0 ~.
	\label{BB}
\end{equation}
This way, following DeWitt\cite{BBboundary} argument, we
'neutralize' the Big Bang singularity by making the origin
quantum mechanically inaccessible to wave packets.

At this stage, while sticking to the full Lorentzian
picture, namely $\Psi=\psi(a)e^{- i\omega \tau}$ even
under the potential barrier, our discussion bifurcates
with respect to the left over boundary condition:

\medskip
\noindent $\bullet$ Following Hartle-Hawking (HH)\cite{HH}
or Linde (L)\cite{L} proposals, where Hermiticity (real
$\omega$) is the name of the game, the naive WKB wave
function under the barrier is given by
\begin{equation}
	\psi_{HH,L}(a_{L}<a<a_{R}) \simeq
	\mp \frac{1}{\sqrt{V}}
	\exp\left[\pm \int_{a_{L}}^{a}\sqrt{V}da'\right] ~,
\end{equation}
respectively.
The corresponding nucleation probability is
\begin{equation}
	{\cal P} \sim
	e^{{\displaystyle \pm 2\int_{a_{L}}^{a_{R}}
	\sqrt{V}da'}} \simeq
	e^{{\displaystyle \pm \frac{4}{3H^{2}}
	(1-\frac{3}{2}H\omega)}} ~.
\end{equation}
The matching at $a=a_{R}$ yields a symmetric
(antisymmetric) combination of equal strength outgoing
and ingoing waves.
The $a=a_{L}$ matching into the Embryonic zone would
contradict the Big-Bang boundary condition eq.(\ref{BB})
unless
\begin{equation}
	\exp\left[2i \left(\int_{0}^{a_{L}}\sqrt{-V}da'
	-\frac{\pi}{4}\right)\right] = \pm 1 ~.
\end{equation}
The result is 'energy' (not to be confused with the
energy $E=0$) quantization.
To be specific, for $\omega H \ll 1$, we invoke
eq.(\ref{well}) and after some algebra derive the
discrete 'energy' spectrums
\begin{equation}
	\omega^{HH,L}_{n} \simeq
	\sqrt{\frac{2}{3}(4n\pm 1)} ~,
\end{equation}
such that $\omega^{L}_{min}=\sqrt{3}\omega^{HH}_{min}$.

Having a non-zero ground state 'energy' $\omega_{min}$
is remarkable.
It is the closest one can get to Einstein limit $\omega=0$.
But what exactly do we mean by a ground state, and why
does the Einstein limit make sense?
A successful (presumably Euclidean) theory of boundary
conditions must explain why is low $\omega$ preferable
to high $\omega$.
\begin{figure}[tbp]
	\begin{center}
		\includegraphics[scale=0.5]{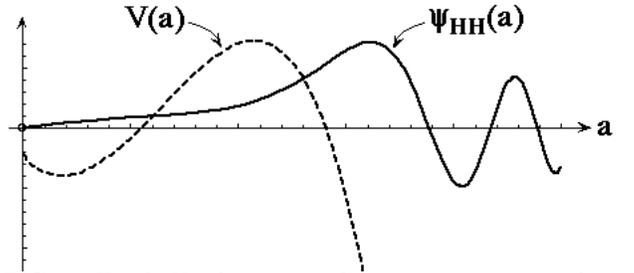}
		\caption{Hartle-Hawking ground state ($n=0$) wave
		function of a brane-like Universe (notice its vanishing
		at the Big Bang). The dashed curve being the underlying
		brane-like Wheeler-DeWitt potential.}
		\label{fig2}
	\end{center}
\end{figure}
\noindent $\bullet$ Vilenkin (V)\cite{V} proposal on the
other hand is characterized by an outgoing wave function
\begin{equation}
	\psi_{V}(a>a_{R}) \sim \frac{1}{\sqrt{-V}}
	\exp\left[i\int_{a_{R}}^{a}\sqrt{-V}da'\right] ~.
\end{equation}
The WKB behavior of the wave function can then be traced
back all the way to the origin where it is supposed to vanish.
The consistency condition then reads
\begin{equation}
	\exp\left[2i\left(\int_{0}^{a_{L}}\sqrt{-V}da'-
	\frac{\pi}{4}\right)\right]
	\simeq \frac{1+4\theta^{2}}{1-4\theta^{2}} ~,
\end{equation}
where $\theta \equiv e^{\int_{a_{L}}^{a_{R}}\sqrt{V}da'}$
is the opacity coefficient.
The latter equation can only be satisfied by a complex 
'energy' $\omega=\tilde{\omega}-\frac{1}{2}i\Gamma$.
It should be noticed how the Hartle-Hawking (Linde)
discrete spectrum, that is $\tilde{\omega}\rightarrow
\omega_{n}^{HH(L)}$ followed by $\Gamma\rightarrow 0$,
is recovered for $\theta \ll 1$ ($\theta \gg 1$).
Altogether, the disintegration of Vilenkin bubble highly
resembles $\alpha$-decay.

Euclidization is next.
In the first glance it may look like the Lorentzian and
the Euclidean regimes share the one and the same Embedding
spacetime, and that  Euclidization can be formulated in
the language of the mini-superspace light-cone.
However, a simple investigation reveals that a closed
Euclidean FRW metric cannot be embedded within a flat
Minkowski spacetime.
It calls for a flat Euclidean background, attainable by
means of Wick rotation $\tau \rightarrow \pm i\tau_{E}$ (with
the corresponding cosmic gauge being $\dot{\tau_{E}}^{2}+
\dot{a_{E}}^{2}=1$).

We are not in a position to tell whether Euclidean
gravity is only a technical tool, serving to explain
certain quantum and/or thermodynamic aspects of the
Lorentzian theory, or perhaps has life of its own.
This way or the other, the emerging picture is of a
Euclidean manifold  sandwiched between two Lorentzian
regimes.
\begin{figure}[tbp]
	\begin{center}
		\includegraphics[scale=0.65]{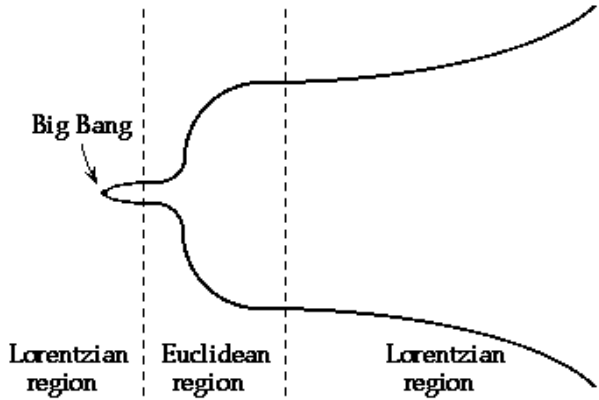}
		\caption{The Euclidean regime sandwiched between
		the Embryonic and the Expanding Lorentzian epochs.}
		\label{fig3}
	\end{center}
\end{figure}
The Euclidean time difference $\delta$ to travel back
and forth the $a_{L}<a<a_{R}$ well of the upside down
potential $-V$ is given by
\begin{equation}
	\delta = 2\int_{a_{L}}^{a_{R}}
	\frac{da}{\sqrt{1-\xi H^{2}a^{2}}} ~,
\end{equation}
and takes the value
\begin{equation}
	\delta \simeq \left\{
	\begin{array}{ccc}
		\frac{\pi}{H}
		\left(1-\frac{2}{\pi}\sqrt{\omega H}\right) &
		~~~\text{if}~ & \omega H \ll 1  ~, \\
		4\sqrt{3}\omega & ~~~\text{if}~  &
		 \omega H =\frac{2}{3\sqrt{3}} ~.
	\end{array}
	\right.
\end{equation}
Recall two relevant facts:
(i) The Euclidean manifold \textit{can} be periodic
in $t_{E}$. The allowed periodicities are restricted, however,
to the sequence $\Delta t_{E} = N\delta$ ($N$ integer).
(ii) At the Euclidean de-Sitter limit, where $\omega
\rightarrow 0$, the Euclidean manifold \textit{must} be
periodic in $t_{E}$ with period $\Delta t_{E}=2\pi H^{-1}$,
as otherwise a conic singularity is present.
Combining these two facts, one can identify $t_{E}$ with
$t_{E}+\Delta t_{E}$ provided
\begin{equation}
	\Delta t_{E} = 2\delta  ~.
\end{equation}
In turn, our bubble Universe is characterized by a
temperature $\displaystyle{T=\frac{1}{\Delta t_{E}}}$
and an entropy $\displaystyle{S=\frac{1}{4\pi}\Delta
t_{E}^{2}}$.

The model discussed here has no pretension to be realistic.
Its objective is primarily pedagogical, to concretely demonstrate
(i) How to overcome the problem (absence) of time in canonical
quantum gravity, and (ii) How to 'neutralize', quantum-mechanically,
the Big-Bang problem at the Lorentzian level.
All this without upsetting the leading wave-function proposals.
It remains to be understood though how to convert the emerging
closed Universe into an open one (following perhaps
Hawking-Turok\cite{open1} prescription), how does inflation enter
the game (presumably along Linde\cite{open2} or Vilenkin\cite{open3}
trails), and whether there exists some leftover experimental crumb.
At any rate, several model independent features, notably the
classically disconnected Embryonic epoch, are to be regarded
as the finger-prints of the underlying theory.
Brane-like Universe gravity constitutes a controlled deviation 
(automatic energy/momentum conservation) from Einstein gravity,
with the latter regarded as the classical ground-state limit.

\acknowledgments
It is our pleasure to thank Professors E. Guendelman and R.
Brustein for valuable discussions and enlightening remarks.

\end{document}